\documentstyle[12pt]{article}
\newcommand{\be}{\begin{equation}}

\newcommand{\ee}{\end{equation}}
\newcommand{\beqa}{\begin{eqnarray}}
\newcommand{\eeqa}{\end{eqnarray}}

\newcommand{\la}{\lambda}

\newcommand{\lb}{\label}
\begin{document}

\title{Hamiltonian Formulation of Jackiw--Pi 3-Dimensional Gauge Theories}
\author{\"{O}mer F. DAYI\thanks{E-mail: dayi@gursey.gov.tr
and dayi@mam.gov.tr.} \\
{\small {\it Feza G\"{u}rsey Institute, }}\\
{\small {\it P.O.Box 6, 81220 \c{C}engelk\"{o}y--Istanbul, Turkey.}}}
\date{}
\maketitle

\begin{abstract}
A 3-dimensional non-abelian gauge theory was proposed by Jackiw and Pi to
create mass for the gauge fields. However,
the quadratic action obtained by switching off the non-abelian interactions
possesses more gauge symmetries
than the original one, causing some difficulties in quantization.
Jackiw and Pi proposed another action
by introducing new fields,
whose gauge symmetries are consistent with
the quadratic part. It is shown that all of these theories have the same
number of physical degrees of freedom in the hamiltonian framework. Hence,
as far as the physical states are considered there is no inconsistency.
Nevertheless, perturbation expansion is still problematic.
To cure this we propose to modify one of the constraints
of the non-abelian theory without altering neither its canonical
hamiltonian nor the number of physical states.

\vspace{3cm}
\end{abstract}

\vspace{2cm}
\noindent
RIBS--PH--TH--9/97

\noindent
hep-th/9711079

\pagebreak

There are some different approaches to generate mass for the gauge fields of
non-abelian gauge theories in 3--dimensions in terms of the gap
equation
$$
\Pi (p^2)|_{p^2=m^2}=m^2, 
$$
where, $\Pi $ is the transverse vacuum polarization tensor. This equation is
studied for different actions up to one loop level leading to some different
results\cite{gen}--\cite{jp2}.
However, considering higher loops seems to be
essential\cite{col},\cite{ebe}.

In the approach of Jackiw and Pi\cite{jp2} one
deals with actions whose quantization in lagrangian
formalism exhibits some uncommon features.
Quadratic action possesses more gauge symmetries than the
non-abelian one. Thus, perturbative expansion of the
latter is not well defined. To cure this 
in Ref.\cite{jp2} another action is proposed to the cost
of introducing new fields. We will show that
considered as  constrained hamiltonian  systems there is
no inconsistency between these actions:
the number of physical states is the same.

Although all of them possess the same number of physical states,
hamiltonian quantization of non-abelian case is still
problematic. To overcome this difficulty,
we propose to modify one of the original constraints of the
non-abelian theory
by making use of gauge fixing conditions of quadratic action,
without altering its canonical hamiltonian and 
number of physical states.

Jackiw and Pi proposed the action\cite{jp2} 
\begin{equation}
\label{fa}S
=\int d^3x \left[- \frac{1}{4}F^{a \mu \nu}F^a_{\mu \nu} -
\frac{1}{4}G^{a \mu \nu}G^a_{\mu \nu} +\frac{m}{2}
\epsilon^{\mu \nu \rho}F_{\mu \nu}^a\phi_\rho^a \right] , 
\end{equation}
where the group index $a=1,\cdots ,N,$ and 
\begin{eqnarray}
F^a_{\mu \nu} & = & \partial_\mu A_\nu^a - \partial_\nu A_\mu^a
+\alpha f^{abc}A_\mu^bA_\nu^c \\
G^a_{\mu \nu} & = & (D_\mu \phi_\nu )^a -(D_\nu \phi_\mu )^a.
\end{eqnarray}
Covariant derivative is given in terms of the structure constants $f^{abc}$
as 
$$
D_\mu^{ab}=\delta^{ab}\partial_\mu +\alpha f^{acb}A_\mu^c . 
$$
(\ref{fa}) is invariant under the gauge transformations 
\begin{equation}
\label{gi1}\delta_1 A_\mu^a =(D_\mu \theta)^a ,\
\delta_1\phi_\mu^a=f^{abc}\phi^b_\mu \theta^c . 
\end{equation}

However, when the coupling is switched off $(\alpha =0)$, 
(\ref{fa}) yields the quadratic action 
\begin{equation}
\label{fr}S_f\equiv S(\alpha =0)
\end{equation}
which is invariant under two different types of abelian gauge
transformations 
\begin{eqnarray}
\delta_{f1} A_\mu =\partial_\mu\theta & , & \delta_{f1} \phi_\mu =0,\\
\delta_{f2} A_\mu =0 & ,
& \delta_{f2} \phi_\mu =\partial_\mu \xi . \label{2g}
\end{eqnarray}
Obviously, (\ref{fa}) is not invariant under the non-abelian generalization
of (\ref{2g}): 
\begin{equation}
\label{gi2}\delta _2A_\mu ^a=0,\ \delta _2\phi _\mu ^a=(D_\mu \xi )^a. 
\end{equation}
Because of this, perturbation quantization of the full (non-abelian) theory
(\ref{fa}) is not straightforward. When (\ref{fa}) is used in Green functions
generating functional or in partition function, there is only need of gauge
fixing terms for its gauge symmetries (\ref{gi1}). But, the propagators
will be calculated in terms of the quadratic action (\ref{fr}), which still
possesses the gauge symmetry (\ref{2g}). i.e. gauge fixing of the
non-abelian action will not be sufficient to eliminate the redundant fields
in (\ref{fr}) which is essential to define finite propagators.

A general quantization procedure of the theories whose gauge
symmetries in the quadratic and the full cases are not consistent is not
available yet.

Jackiw and Pi proposed to enlarge the space of states by introducing the new
fields $\rho^a$ and to deal with the action 
\begin{equation}
\label{sg}S_g=\int d^3x \left[- \frac{1}{4}F^{a \mu \nu}F^a_{\mu \nu} - 
\frac{1}{4}(G^{a \mu \nu}+f^{abc}F^{b \mu \nu}\rho^c ) (G^a_{\mu \nu}
+f^{ab^\prime c^\prime} F^{b^\prime}_{\mu \nu}\rho^{c^\prime} ) +\frac{m}{2}
\epsilon^{\mu \nu \rho}F_{\mu \nu}^a\phi_\rho^a \right] , 
\end{equation}
which is invariant under both of the gauge transformations (\ref{gi1})
and (\ref{gi2}) supplemented by 
\begin{equation}
\label{gir}\delta_1 \rho^a = f^{abc}\rho^b\theta^c,\ \delta_2\rho^a=-\xi^a . 
\end{equation}

We would like to analyze the above mentioned actions in terms of 
the hamiltonian methods.

Let us first deal with the quadratic case (\ref{fr}). By using the
definition of canonical momenta 
\begin{equation}
\pi _\mu ^a\equiv \frac{\delta S}{\delta A^{a\mu }};\ P_\mu^a
\equiv \frac{\delta S}{\delta \phi ^{a\mu }}, 
\end{equation}
we obtain the primary constraints 
\begin{equation}
\label{pc}\pi _0^a=0,\ P_0^a=0, 
\end{equation}
and the canonical hamiltonian 
\begin{eqnarray}
H_f & =\int d^2x & {\Large [} \frac{1}{2} (\pi_i^a)^2 +\frac{1}{2}(P_i^a)^2-
m\epsilon_{ij} \pi^{ai}\phi^{aj}
+\frac{1}{2}m^2(\phi^a_i)^2 +
\frac{1}{2}
(\partial_iA_j^a -\partial_jA_i^a)^2 \nonumber \\
& &+\frac{1}{2} (\partial_i\phi_j^a-\partial_j\phi_i^a)^2
-A_0^a\psi_{f1}^a
-\phi_0^a\psi_{f2}^a  {\Large ]} , \label{hf}
\end{eqnarray}
where we use the  metric $\eta _{\mu \nu }={\rm diag}(-1,1,1)$ and the
definitions 
\begin{eqnarray}
\psi_{f1}^a & \equiv  & \partial^i \pi_i^a ,\\
\psi_{f2}^a & \equiv  & \partial_i   P_i^a +m\epsilon^{ij}\partial_iA^a_j.
\end{eqnarray}
Obviously, here the group index $a$ is for $N$ copies of $U(1).$

Time evolution of classical observables will be given by the extended
hamiltonian 
$$
H^\prime_f=H_f+\int d^2x\left[ \kappa _1^a\pi _0^a+\kappa _2^aP_0^a\right] . 
$$

The primary constraints (\ref{pc}) should be preserved in
time\footnote{We deal with the equal time Poisson brackets 
$$
\{ {\cal O}(x),{\cal K}(y) \} = \int d^2z \left[
\frac{\delta {\cal O}(x)}{\delta p^I(z)}
\frac{\delta {\cal K}(y)}{\delta q_I(z)}
- \frac{\delta {\cal O}(x)}{\delta q_I(z)}
\frac{\delta {\cal K}(y)}{\delta p^I(z)} \right]
$$
where ${\cal O},\ {\cal K}$ are some classical observables and $q_I=
(A_\mu,\phi_\mu),\ p^I= (\pi^\mu ,P^\mu).$} on the constraint surface
defined by vanishing of the related constraints ( denoted by $\approx $): 
$$
\dot{\pi}_0^a(x)=\{\pi _0^a(x),
H^\prime_f\}\approx 0;\  \dot{P}_0^a(x)=
\{P_0^a(x),H^\prime_f\}\approx 0.
$$
These lead to the secondary constraints 
\begin{equation}
\label{ao1}\psi _{f1}^a(x)=0;\ \psi _{f2}^a(x)=0,
\end{equation}
which are conserved in time: 
\begin{eqnarray}
\{\psi_{f1}^a(x),H^\prime_f\} & \approx & 0 ,\\
\{\psi_{f2}^a(x),H^\prime_f\} & \approx & 0.
\end{eqnarray}
Thus, there is no more constraint. Moreover, all of the
constraints (\ref{pc})
and (\ref{ao1}) give vanishing Poisson brackets among themselves.

In the non-abelian case (\ref{fa}) the canonical hamiltonian is 
\begin{eqnarray}
H & =\int d^2x & {\Large [} \frac{1}{2} (\pi_i^a)^2 +\frac{1}{2}(P_i^a)^2-
m\epsilon_{ij} \pi^{ai}\phi^{aj}
+\frac{1}{2}m^2(\phi^a_i)^2 +
\frac{1}{2}F^{a ij}F^a_{ij} \nonumber \\
& &+ \frac{1}{2}G^{a ij}G^a_{ij}
-A_0^a\tilde{\psi}_1^a
-\phi_0^a\tilde{\psi}_2^a  {\Large ]},          \label{ch}
\end{eqnarray}
where 
\begin{eqnarray}
\tilde{\psi}_1^a & \equiv  & (D_i \pi^i)^a +\alpha f^{abc}\phi_i^bP^{ic} ;\\
\tilde{\psi}_2^a & \equiv  & (D_i   P^i)^a +\frac{m}{2}\epsilon^{ij}F^a_{ij}.
\end{eqnarray}

Primary constraints are still given by (\ref{pc}). So that, the extended
hamiltonian is 
$$
H^\prime =H+\int d^2x \left[ \kappa_1^a \pi_0^a+\kappa_2^aP_0^a \right] . 
$$

Conservation of the primary constraints (\ref{pc}) in time
leads to the secondary constraints 
\begin{equation}
\tilde{\psi}_1^a=0;\ \tilde{\psi}_2^a=0. 
\end{equation}

Poisson brackets satisfied by the secondary constraints $\tilde{\psi}_1^a,\ 
\tilde{\psi}_2^a$ are 
\begin{eqnarray}
\{\tilde{\psi}_1^a(x),\tilde{\psi}_1^b(y)\} & = &
-\alpha f^{abc}\tilde{\psi}_1^c(x) \delta (x-y), \\
\{\tilde{\psi}_1^a(x),\tilde{\psi}_2^b(y)\} & = &
-\alpha f^{abc}\tilde{\psi}_2^c(x) \delta (x-y), \\
\{\tilde{\psi}_2^a(x),\tilde{\psi}_2^b(y)\} & = & 0 .
\end{eqnarray}
By making use of these relations one can show that 
\begin{eqnarray}
\{\tilde{\psi}_1^a(x),H^\prime\} & \approx & 0 ,\\
\{\tilde{\psi}_2^a(x),H^\prime\} & \approx  & - \tilde{\psi}_3^a(x),
\end{eqnarray}
where 
\begin{equation}
\tilde{\psi}_3^a =\alpha f^{abc}[ F_{ij}^b G^{cji} -P_i^b ( \pi^{ic} -m
\epsilon^{ij} \phi^c_j)]. 
\end{equation}
Hence, conservation of the secondary constraints yields the new constraints 
$$
\tilde{\psi}_3^a=0. 
$$

Obviously, $\pi_0^a $ and $P_0^a$ give vanishing Poisson brackets with the
other constraints $\tilde{\psi}_{1,2,3}^a.$ However, 
\begin{eqnarray}
\{\tilde{\psi}_1^a(x),\tilde{\psi}_3^b(y)\} & = &
-\alpha f^{abc}\tilde{\psi}_3^c(x) \delta (x-y), \\
\{\tilde{\psi}_2^a(x),\tilde{\psi}_3^b(y)\} & = &
\alpha^2 f^{acd}f^{b d^\prime c} [ P_i^dP_i^{d^\prime}
+F_{ij}^dF_{ij}^{d^\prime} ]\delta (x-y). \label{nv}
\end{eqnarray}
Thus, by denoting the canonical hamiltonian
(extended hamiltonian evaluated on the
constraint surface)  by $H_0,$
the condition that the constraints $\tilde{\psi}_3^a(x)$
should be conserved in time
\begin{equation}
\label{aa}
\{\tilde{\psi}_3^a(x),H^\prime\} =\{\tilde{\psi}_3^a(x),H_0\} -\int
d^2y \phi_0^b(y) \{\tilde{\psi}_3^a(x),\tilde{\psi}_2^b(y)\} \approx 0, 
\end{equation}
will yield a solution for $\phi_0^a,$ as far as we exclude the
configurations which make the right hand side of (\ref{nv}) vanishing. Thus,
by accepting that the right hand side of (\ref{nv}) is non-vanishing we
conclude that there is no more constraint. Although (\ref{aa}) may lead to a
non-local $\phi_0^a(x),$ for the functional integrals the relevant
hamiltonian is the one evaluated on the constraint surface where the term
including $\phi_0^a(x)$ is absent.

For the action $S_g$ (\ref{sg}) primary constraints are 
\begin{equation}
\label{prg}\pi _0^a=0;\ P_0^a=0;\ \lambda ^a\equiv \frac{\delta S_g}{\delta
\rho ^a(x)}=0. 
\end{equation}
Hence, the extended hamiltonian reads 
\begin{eqnarray}
H_g & =H^\prime +\int d^2x {\Large [} &
\alpha f^{abc}\rho^a (m\epsilon^{ij} P_i^b\phi_j^c -
P_i^b\pi^{ic} +G_{ij}^bF^{c ij}) \nonumber \\
& & +\frac{\alpha^2}{2}f^{adc}f^{c b  d^\prime }\rho^a\rho^b
(P_i^dP^{id^\prime } +F_{ij}^d F^{d^\prime ij})
+ \kappa_3^a \lambda^a {\Large ]}.
\end{eqnarray}

Vanishing of Poisson brackets (now including also derivatives with respect
to $\rho $ and $\lambda $) of the primary constraints (\ref{prg}) with the
extended hamiltonian $H_g$ will yield as before 
\begin{equation}
\tilde \psi _1^a(x)=0;\ \tilde \psi _2^a(x)=0. 
\end{equation}
Moreover, there are the following secondary constraints 
\begin{equation}
\psi _3^a\equiv \tilde \psi _3^a+\alpha ^2f^{adc}f^{cbd^{\prime }}\rho
^b(P_i^dP^{id^{\prime }}+F_{ij}^dF^{d^{\prime }ij})=0. 
\end{equation}
Because of the fact that 
\begin{equation}
\{\psi _3^a(x),\lambda ^b(y)\}=-\alpha ^2f^{adc}f^{cbd^{\prime
}}(P_i^dP_i^{d^{\prime }}+F_{ij}^dF^{d^{\prime }ij})\delta (x-y), 
\end{equation}
which is assumed to be non-vanishing, constraints are terminated: 
$$
\{H_g,\psi _3^a(x)\}\approx 0 
$$
will be satisfied by choosing $\kappa _3^a(x)$ appropriately.

Let us deal with the following linear combination of the constraints 
\begin{eqnarray}
\psi^a_1(x) & \equiv  & \tilde{\psi}_1^a(x) +\alpha f^{abc}
\rho^b(x) \lambda^c(x),\\
\psi^a_2(x) & \equiv  & \tilde{\psi}_2^a(x) + \lambda^a(x).
\end{eqnarray}
which satisfy the Poisson bracket relations 
\begin{eqnarray}
\{ {\psi}_1^a(x), {\psi}_1^b(y)\} & = &
-\alpha f^{abc} {\psi}_1^c(x) \delta (x-y), \\
\{ {\psi}_1^a(x), {\psi}_2^b(y)\} & = &
-\alpha f^{abc} {\psi}_2^c(x) \delta (x-y), \\
\{ {\psi}_1^a(x), {\psi}_3^b(y)\} & = &
-\alpha f^{abc} {\psi}_3^c(x) \delta (x-y), \\
\{\psi_1^a(x),\lambda^b(y) \} & = & -f^{abc}\lambda_c \delta (x-y),\\
\{ {\psi}_2^a(x), {\psi}_3^b(y)\} & = & 0, \\
\{ {\psi}_2^a(x), {\psi}_2^b(y)\} & = & 0 .
\end{eqnarray}
One can show that the new constraints possess consistent equations of
motion: 
\begin{eqnarray}
\{\psi_1^a(x),H_g\} & \approx & 0 ,\\
\{\psi_2^a(x),H_g\} & \approx  & 0,
\end{eqnarray}
where, the constraint surface is defined in terms of the new set of
constraints.

Let us classify the constraints a l\'a Dirac\cite{di} to find  number of
physical degrees of freedom (at least in reduced phase space method). These
are listed below for the three cases: {\it quadratic} given by (\ref{fr}), 
{\it non-abelian} given by (\ref{fa}), {\it enlarged} given by (\ref{sg}). 
\begin{flushleft}
\begin{tabular}{|l|c|c|c|}   \hline
 & {\it quadratic} & {\it non-abelian} & {\it enlarged} \\ \hline
First class & & &  \\
 constraints &
$\pi_0^a,\ P_0^a,\ \psi_{f1}^a,\ \psi_{f2}^a. $   &
 $\pi_0^a,\ P_0^a,\ \tilde{\psi}_{1}^a. $&
$\pi_0^a,\ P_0^a,\ \psi_{1}^a,\ \psi_{2}^a. $ \\ \hline
Second class & & &  \\
constraints & - &
$ \tilde{\psi}_{2}^a,\ \tilde{\psi}_{3}^a.$ &
$ \psi_3^a,\ \lambda^a $\\ \hline
\# physical phase & & &  \\
space variables
& $12N-8N=4N $ & $12N-6N-2N=4N$ & $14N-8N-2N=4N $\\ \hline
\end{tabular}
\end{flushleft}

Hence, as far as the physical states are concerned there is no inconsistency
between the three cases considered. There is no negative norm state in any
of the Hilbert spaces in the quantum case.

Let us discuss the $\alpha =0$ limit of the non-abelian and the enlarged
cases.

\noindent
{\it i) non-abelian:} $H|_{\alpha =0}=H_f,$ $\tilde \psi _1|_{\alpha
=0}=\psi _{f1},$ $\tilde \psi _2|_{\alpha =0}=\psi _{f2}$ and $\tilde \psi
_3|_{\alpha =0}=0.$ So that, the second class constraints $\tilde \psi _2$
become first class. In principle, by making appropriate changes in the
related hamiltonian, one can consider $\tilde \psi _2$ as first class
constraints and $\tilde \psi _3$ as their gauge fixing (subsidiary)
conditions. However, these will cease to be gauge fixing conditions for the
quadratic case. In fact, this explains how the inconsistency of the gauge
symmetries of the two cases arises in the lagrangian formalism, although
the number of physical degrees of freedom is the same.

\noindent
{\it ii) enlarged:} $H_g|_{\alpha =0}=H_f,$ $\psi _1|_{\alpha =0}=\psi
_{f1}, $ $\psi _2|_{\alpha =0}=\psi _{f2}+\lambda $ and $\psi _3|_{\alpha
=0}=0.$ Now, $\lambda ^a=0$ are first class. In this case one can adopt the
gauge fixing conditions $\chi _2^a=0$ corresponding to the first class
constraints $\psi _2$, yielding gauge fixing conditions for $\psi _{f2}$ in
the $\alpha =0$ limit. However, there are some other problems in the
perturbation expansions:

Let us consider the functional integral 
\begin{equation}
\label{fin}{\cal Z} =\int d\mu [p,q] \exp\int d^3x (p_A \dot{q}_A -{\cal H}
), 
\end{equation}
where $q_A$ indicate $A_\mu^a,\ \phi_\mu^a,$ in the quadratic and
non-abelian cases and also $\rho^a$ in the enlarged case and ${\cal H}$ is
the related hamiltonian density. Let us separate the measure $d\mu [p,q]$ as 
$$
d\mu [p,q] =\mu_0 \mu_e dp_Adq_A, 
$$
where $\mu_0$ is the part related to the first class constraints $\pi_0,\
P_0,$ and one of $\psi_{f1},\ \tilde{\psi}_1,\ \psi_1$ depending on the
action and their subsidiary conditions, which do not cause any difficulty.

For the enlarged case the other part of the measure is 
\begin{equation}
\mu_e =\delta (\psi_2) \delta (\chi_2) \delta (\psi_3) \delta (\lambda )
\det \left[ \alpha^2 f^{adc}f^{c b d^\prime } (P_i^dP_i^{d^\prime }
+F_{ij}^d F^{d^\prime ij}) \right] \det \{\psi_2 ,\chi_2\}. 
\end{equation}
Thus, the $\alpha =0$ limit is not well defined for the functional integral.
By integrating over $\rho$ the measure will possess the term $\det^{1/2}
\left[ \alpha^2 f^{adc}f^{c b d^\prime } (P_i^dP_i^{d^\prime } +F_{ij}^d
F^{d^\prime ij}) \right]$ as it was announced in \cite{jp2}.

If one would like to obtain the non-abelian case from the enlarged one,
gauge fixing conditions can be chosen $\chi _2^a=\rho ^a.$ After integrating
over $\rho $ and $\lambda :$ $H_g \rightarrow H$ and the related part of the
measure will yield

$$
\mu _e=\delta (\tilde \psi _2)\delta (\tilde \psi _3)\det [\delta^{ab}
\delta (0)]\det \left[ \alpha ^2f^{adc}f^{cbd^{\prime }}(P_i^dP_i^{d^{\prime
}}+F_{ij}^dF^{d^{\prime }ij})\right] . 
$$
The term $\det [\delta (0)]$ can be absorbed by the normalization of the
partition function.

A way to cure the non-abelian theory would be to consider a combination of
the second class constraints as $\Sigma^a (x)=\kappa_2^{ab}(x)
\tilde{\psi}_2^b(x) + \kappa_3^{ab}(x) \tilde{\psi}_3^b(x), $ where $
\kappa_{2,3}^{ab}(x) $ are defined to satisfy $\{\Sigma^a(x),\Sigma^b(y)\}
\approx 0 ,$ where the constraint surface is defined by vanishing of the
other constraints and $\Sigma^a(x)=0.$ Moreover, we should define a new
hamiltonian $\tilde{H}$ satisfying $\{\Sigma^a(x),\tilde{H}\}\approx 0 $ and 
$\tilde{H}\approx H.$ Once this is achieved one can adopt gauge fixing
conditions which do not vanish for $\alpha =0.$ However, this may force to
introduce some undesired non-local terms.

We propose another way of resolving the problem.
We reinterpret the coordinate fields of the
primary constraints
$P_0=\pi_0=0$ as Lagrange
multipliers:
\be
\lb{rep}
A_0^a,\phi_0^a \rightarrow \lambda_1^a,\lambda_2^a,
\ee
and do not consider zero components of the fields any more.

In the quadratic case we may introduce the gauge fixing
conditions
\[
\chi_{f1}^a (x)= 0,\ \chi_{f2}^a (x) =0,
\]
satisfying
\be
\lb{fgf}
\det \{ \psi_{f1}^a(x), \chi_{f1}^b(y) \} \neq 0,\,
\det \{ \psi_{f2}^a(x), \chi_{f2}^b (y)\} \neq 0.
\ee
The appropriate hamiltonian is given by $H_f$ (\ref{hf}),
after the replacement
(\ref{rep}), where $\lambda_1^a,\ \lambda_2^a$
are defined such that
\[
\{H_f,\chi_{f1}^a(x)\}\approx 0 ,\  \{H_f,\chi_{f2}^a(x)\}\approx 0.
\]
Here the constraint surface is defined in terms of the
original constraints and the gauge fixing conditions.

In the non-abelian case we need to introduce gauge fixing
conditions for $\tilde{\psi}_1^a(x) :$
\[
\tilde{\chi}_1^a(x)=0,
\]
satisfying
\[
\det \{\tilde{\psi}_1^a(x),\tilde{\chi}_1^b(y)\}\neq 0.
\]
Moreover,
\[
\{H^\prime,\tilde{\chi}_1^a(x)\} \approx 0,
\]
will lead an equation for the related Lagrange multipliers
$\lambda_1^a(x).$ Obviously, $\lambda_2^a(x)$ have already been fixed
by the condition (\ref{aa}).

Instead of the original non-abelian one,
we propose to deal with the constrained hamiltonian system
given by the following set of constraints
\be
\lb{nsc}
\Phi_A\equiv (
\tilde{\psi}_1,\ \tilde{\chi}_1,\ \tilde{\psi}_2,\ \tilde{\chi}_2 )=0,
\ee
where the modified constraints are
\[
\tilde{\chi}_2\equiv \chi_{f2} + \tilde{\psi}_3.
\]
However, the hamiltonian is still given by
(\ref{ch}), with the replacement (\ref{rep}). Because
of being second class $\Phi_A$ satisfy 
\[
\det \{\Phi_A(x),\Phi_B(y)\} \neq 0.
\]
So that, the
Lagrange multipliers $\lambda_1^a(x),\ \lambda_2^a(x)$
are given as solutions of the equations
\[
\{H, \Phi_A(x)\}\approx 0.
\]
Now the limit $\alpha =0 $ is well defined and the number
of the physical states are unaltered.
Obviously, the  mass induced by this theory
should be calculated to see if  it is satisfactory. However,
it is out of the scope of this paper.

Arnowitt and Deser studied a theory similar to (\ref{fa})
in 4--dimensions\cite{ad} (obviously the last term in the action
is absent) exhibiting the same features of gauge transformations.
Unfortunately, on the contrary, in the case of
Ref. \cite{ad}  hamiltonian approach will yield inconsistency
between the numbers of physical states of quadratic and
non-abelian theories\footnote{I would like to thank
R. Jackiw and S-Y. Pi for asking me
to comment on this point.}: in 3--dimensions one of the gauge
symmetries of the quadratic action is preserved after introducing
the non-abelian terms. This manifest itself as secondary
first class constraint, namely $\tilde{\psi}_1,$ which comes
from the condition $\dot{P}_0=0.$ The other primary constraint
$\pi_0=0$ leads to  $\tilde{\psi}_2.$ The latter is related to the
gauge symmetry of the quadratic action  which is broken
in the non-abelian case and it leads to the constraint
$\tilde{\psi}_3=0.$ So that, the number of the physical states
of the abelian and the non-abelian cases is the same.
However, in the 4--dimensional analog all of the gauge
symmetries of the quadratic action are broken after
introducing the non-abelian terms. Thus, none of
the secondary constraints is first class.
Preserving them in time does not lead to
any other constraint but dictate  form of the
Lagrange multipliers $\la_1,\ \la_2.$

\pagebreak



\begin{thebibliography}{9}
\bibitem{gen}  G. Alexanian and V.P. Nair, Phys. Lett. B 352 (1995) 435;\\
W. Buchm\"uller and O. Philipsen, Nucl. Phys. B 443 (1995) 47; Phys. Lett. B
397 (1997) 112.;\\ D. Comelli and M. Pietroni, DFPD 97/TH/37,
hep-ph/9708489. 

\bibitem{jp1}  R. Jackiw and S-Y. Pi, Phys. Lett. B 368 (1996) 131.

\bibitem{jp2}  R. Jackiw and S-Y. Pi, Phys. Lett. B 403 (1996) 297.

\bibitem{col}  J.M. Cornwall, UCLA/97/TEP/12, hep-th/9710128.

\bibitem{ebe}  F. Eberlein, ISSS 0418-9833, hep-th/9804460

\bibitem{di}  P.A.M. Dirac, Lectures in quantum mechanics (Academic Press,
New York, 1965).

\bibitem{ad}R. Arnowitt and S. Deser, Nucl. Phys. 49 (1963) 133.

\end{thebibliography}
\end{document}